\documentclass{article}
\usepackage{spconf,amsmath,epsfig}
\usepackage{soul}
\usepackage{amssymb}
\usepackage{float}
\usepackage{times}
\usepackage{multirow}
\usepackage{graphicx,epsfig,graphics}

\def\apjl{\ {ApJ}}                
\def\apjs{\ {ApJS}}               
\def\ao{\ {Appl.~Opt.}}           
\def\aap{\ {A\&A}}                

\title{Clustered Radio Interferometric Calibration}
\name{S. Kazemi$^1$, S. Yatawatta$^{1,2}$, S. Zaroubi$^{1,3}$\thanks{The authors acknowledge the Netherlands Foundation
for Scientific research (NWO) grant 436040 for support and SZ thanks the Lady Davis grant.}}
\address{$^1$ Kapteyn Astronomical Institute, Univ. of Groningen, The Netherlands\\
$^2$ ASTRON, The Netherlands\\
$^3$ Technion, Physics Department, Haifa 32000, Israel}
\begin{document}
%
\maketitle
\begin{abstract}
This paper introduces an amendment to radio interferometric calibration of sources below the noise level. The main idea is to employ the information of the stronger sources' measured signals as a plug-in criterion to solve for the weaker ones. For this purpose, we construct a number of source clusters, with centroids mainly near the strongest sources, assuming that the signals of the sources belonging to a single cluster are corrupted by almost the same errors. Due to this characteristic of clusters, each cluster is calibrated as a single source, using all the coherencies of its sources simultaneously. The obtained solutions for every cluster are assigned to all the cluster's sources. An illustrative example reveals the superiority of this calibration compared to the un-clustered calibration.
\end{abstract}
\section{Introduction}
\label{sec:intro}
Calibration of radio synthesis arrays refers to the estimation and reduction of errors introduced by the atmosphere and the incorporated instruments, before imaging. It is the most crucial task in order to achieve the interferometer's desired precision and sensitivity.
Early radio astronomy used external (classical) calibration which is based on estimating the instrument unknown parameters by a celestial radio source with known properties. The external calibration is then improved by self-calibration \cite{selfcal}, which utilizes only the observed data for estimating both the source and instrumental unknowns. 

Although the calibration of radio telescopes highly benefits from various self-calibration techniques, its performance in interferometric source subtraction is still limited to sources that have a high enough Signal to Noise Ratio (SNR) to be distinguished from the background noise \cite{Bernardi, Liu, Pindor}. The novelty of this paper is that the presented method has a high performance in source calibration below the noise level, utilizing the strongest sources' signals. The implementation of such a calibration, termed as "clustered calibration", is performed by clustering the sources during the calibration process. The clustered calibration improves the information used for calculating solutions by incorporating the total of signals observed at each cluster instead of each individual source's signal. Thus, in the case of calibrating the low signals of very weak sources, it provides a considerably better result compared with the un-clustered calibration.

Clustering, from the data mining point of view, can be defined as the gathering of similar data points together into groups. An overview of different clustering methods is given by \cite{Clusteringbook}. We intend to cluster radio sources in the way that visibilities of sources belonging to a single cluster are affected by almost the same errors, and subsequently could share the same calibration's solutions. This assumption is valid when the clusters' angular diameters are small enough so that the variation in their actual solutions is negligible. After arranging the cluster's centroids mainly near the strongest sources, we calibrate considering every cluster as a single source, and assign the obtained solutions to all the cluster's sources. 

We present the data model of clustered calibration and apply two clustering methods: (i) weighted K-means \cite{kmeans, weightedk-means} and (ii) divisive hierarchical clustering \cite{Hierarchicalclustering}, to cluster the sky's sources. In an illustrative example, we demonstrate the superiority of the clustered calibration compared to the un-clustered calibration, in solving for the sources bellow the noise level, using data observed by the LOw Frequency ARray (LOFAR) synthesis radio telescope \footnote{http://www.lofar.org}.  

The following notations are used in this paper: Bold, lower case letters refer to column vectors, e.g., {\bf y}. Upper case bold letters refer to matrices, e.g., {\bf C}.  All parameters are complex numbers, unless stated otherwise. The transpose and conjugation of a matrix are presented by $(.)^T$ and $(.)^*$, respectively. The matrix Kronecker product, the set membership, the empty set, and the union operator are denoted by $\otimes$, $\in$, $\varnothing$, and $\cup$, respectively.

\section{Clustered Calibration Data Model}
\label{Clustered Calibration Data Model}
In this section, we briefly describe the data model of the radio interferometric clustered calibration. For more details on radio interferometry the reader is referred to \cite{A.R.1} and for the data model of radio interferometric calibration to \cite{J.P.1, J.P.2, Panosthesis}.

Consider an interferometric array consisting of $N$ receivers, each with two orthogonal polarized feeds $X$ and $Y$. The induced voltages at the $p$-th receiver's feeds, ${\tilde{\bf{v}}}_{pi}=[{\tilde{v}_{Xpi}}\ {\tilde{v}_{Ypi}}]^T$, due to the polarized waves radiated by the $i$-th source, ${\bf{e}}_i=[e_{Xi}\ e_{Yi}]^T$, is given by 
\begin{equation}
{\tilde{\bf{v}}}_{pi}={\tilde{\bf{J}}}_{pi}{{\bf{e}}}_{i}.\label{s1}
\end{equation}
In Eq. (\ref{s1}), ${\tilde{\bf{J}}}_{pi}$ represents the $2\times 2$ Jones matrix \cite{J.P.1}, corresponding to corruptions of ${\bf{e}}_i$ signal at receiver $p$, which is a product of different Jones matrices as \cite{Panosthesis}
\begin{equation}
{\tilde{\bf{J}}}_{pi}\equiv {\bf{G}}_p{\bf{E}}_{pi}{\bf{Z}}_{pi}{\bf{F}}_{pi}{\bf{K}}_{pi}.\label{i3}
\end{equation} 
In Eq. (\ref{i3}), ${\bf{K}}_{pi}$, ${\bf{E}}_{pi}$, ${\bf{Z}}_{pi}$, and ${\bf{F}}_{pi}$ are the Fourier transform, antenna's voltage pattern, ionospheric phase fluctuation, and Faraday Rotation matrices corresponding to the $i$-th source's direction and receiver $p$'s location on the earth, respectively, and ${\bf{G}}_p$ is the receiver $p$'s clock phase and electronic gain. 

We assume that the total of signals seen at each receiver is a superposition of the $K$ sources' corrupted signals, plus the receiver's thermal noise. Note that the multitude of the ignored fainter sources also contribute to the noise. After correcting the geometric delays corresponding to receivers' locations on the earth, we correlate the collected signals at every pair of receivers to obtain visibilities \cite{J.P.1}. Since the complex gain of Jones matrix ${\bf{G}}$ does not depend on the source direction, it is initially calculated at every receiver and then the visibilities are corrected for it. Stacking up all the corrected visibilities in vector ${\bf y}$, we arrive to the general data model of \cite{S.2, S.K} as
\begin{equation}
{\bf y}=\sum_{i=1}^K {\bf s}_i+{\bf n}.\label{i1}
\end{equation}
In Eq. (\ref{i1}), ${\bf n}$ is the additive noise vector, normally assumed to be Gaussian white noise. The nonlinear function ${\bf s}_i$ shows the contribution of the $i$-th source in the observation:
\begin{equation*}
{\bf s}_i\equiv\left[\begin{array}{c}
{\bf J}^*_{2i}\otimes{\bf J}_{1i}\mbox{vec}({\bf C}_{\{12\}i})\\[-1mm]
\vdots\\[-1mm]
{\bf J}^*_{Ni}\otimes{\bf J}_{(N-1)i}\mbox{vec}({\bf C}_{\{(N-1)N\}i})\end{array}\right],
\end{equation*}
where 
\begin{equation}
{\bf J}_{pi}\equiv{\bf{E}}_{pi}{\bf{Z}}_{pi}{\bf{F}}_{pi},\quad {\bf C}_{\{pq\}i}\equiv{\bf K}_{\{pq\}i}{\bf C}_{i},\label{i33}
\end{equation}
${\bf C}_{i}$ is the source's coherency matrix \cite{bornwolf, J.P.1}, and the scalar Jones matrix ${\bf K}_{\{pq\}i}$ corresponds to Fourier transform between the source's direction and the baseline $pq$.

The calibration is essentially an estimation of the ${\bf{J}}$ Jones matrix, and the removal of the $K$ brightest sources. However, in practice, the ${\bf{E}}$, ${\bf{Z}}$, and ${\bf{F}}$ Jones matrices obtained for nearby directions and for a given receiver are almost the same. Thus, for every receiver $p$, if the $i$-th and $j$-th sources have a small angular separation from each other we have
\begin{equation}
{\bf J}_{pi}\approx {\bf J}_{pj}.\label{i4}
\end{equation}  
Eq. (\ref{i4}) is the underlying assumption for clustered calibration and it tells us that the Fourier transform ${\bf{K}}$ Jones matrix is the only Jones matrix which should be calculated individually for all directions. 

Assume that we have $Q$ ($Q\ll K$) source clusters $L_i$, for $i\in\{1,\ldots,Q\}$, with small enough angular diameters. Based on  Eq. (\ref{i4}), for every cluster $L_i$, we  define 
\begin{equation}
{{\bf{C'}}}_{\{pq\}i}\equiv\sum_{l\in L_i} {\bf C}_{\{pq\}l},\label{i5}
\end{equation}
and by substituting this new definition in Eq. (\ref{i1}), we formulate the clustered calibration data model, where the index $i$ is over the clusters and not separate sources. Various techniques can be used to solve this non-linear data model and one of the more popular of them is the Least Squares (LS) optimization algorithm which is discussed along other methods in \cite{S.2, S.K}.

\section{Clustering of radio sources}
\label{Clustering radio sources}
Suppose that the $K$ sources, $x_1,\ldots,x_K$ have equatorial coordinates (Right Ascension $\alpha$, Declination $\delta$) equal to $(\alpha_1,\delta_1),\ldots,(\alpha_K,\delta_K)$. The aim is to find the optimum $Q$ clusters so that the objective function $f=\sum_{q=1}^Q D(L_q)$ is minimized. $D(L_q)$ is the angular diameter of cluster $L_q$, for $q\in\{1,\ldots,Q\}$, defined as
\hspace*{-6mm}\begin{equation}
D(L_q)\equiv \mbox{max}\ \{d(x_i,x_j)|x_i,x_j\in L_q\},
\end{equation}
and $d(.,.)$ is the angular separation between any two points on the celestial sphere. Having two radio sources $a$ and $b$ with equatorial coordinates $(\alpha_a,\delta_a)$ and $(\alpha_b,\delta_b)$, respectively, the angular separation $d(a,b)$, in radians, is obtained by
\begin{equation}
\mbox{tan}^{-1}\frac{\sqrt{\mbox{cos}^2\delta_b\mbox{sin}^2\Delta\alpha+[\mbox{cos}\delta_a\mbox{sin}\delta_b-\mbox{sin}\delta_a\mbox{cos}\delta_b\mbox{cos}\Delta\alpha]^2}}{\mbox{sin}\delta_a\mbox{sin}\delta_b+\mbox{cos}\delta_a\mbox{cos}\delta_b\mbox{cos}\Delta\alpha},
\end{equation}\label{sp1}
where $\Delta\alpha=\alpha_b-\alpha_a$. 

To get the most information from the strongest observed signals in calibration, the centroids of the clusters should lean towards the brightest sources. Therefore, for defining the centroids, we associate a weight to the $i$-th source, $i\in\{1,\ldots,K\}$, as
\begin{equation}
w_i=w(x_i)\equiv \frac{I_i}{I^*},
\end{equation}\label{Eq.2}
where $I_i$ is the source's intensity and $I^*=\mbox{min}\ \{I_1,\ldots,I_K\}$.

We cluster radio sources using weighted K-means and divisive hierarchical clustering algorithms. Since the source clustering for calibration is performed offline, its computational complexity is negligible compared with the calibration procedure itself. Both of the clustering methods are hard clustering techniques which divide data to distinct clusters. However, we expect more accurate results using fuzzy (soft) clustering, which constructs overlapping clusters with uncertain boundaries. Application and performance of this type of clustering will be explored in future work.

\subsection{Weighted K-means clustering algorithm}
\label{Weighted K-means clustering algorithm}
{\bf Step1.} Select the $Q$ brightest sources, $x_{1^*},\ldots,x_{Q^*}$, and initialize the centroids of $Q$ clusters by their locations as
\begin{equation}
c_q\equiv [\alpha_{q^*},\delta_{q^*}],\quad \mbox{for}\ q\in\{1,\ldots,Q\},\ q^*\in\{1^*,\ldots,Q^*\}.
\end{equation}
{\bf Step2.} Assign each source to the cluster with the closest centroid, defining the membership function
\begin{equation*}
m_{L_q}(x_i)=\begin{cases} 1, & \mbox{if }\  d(x_i,c_q)=\mbox{min}\{d(x_i,c_j)|j=1,\ldots,Q\}  \\ 0,  & \mbox{Otherwise } \end{cases}
\end{equation*}
{\bf Step3.} Update the centroids by
\begin{equation}
c_q=\frac{\sum_{i=1}^Km_{L_q}(x_i)\ w_ix_i}{\sum_{i=1}^Km_{L_q}(x_i)\ w_i},\quad \mbox{for}\ q\in\{1,\ldots,Q\}.
\end{equation}
Repeat steps 2 and 3 until there are no reassignments of sources to clusters.

\subsection{Divisive hierarchical clustering algorithm}
\label{Divisive hierarchical clustering algorithm}
{\bf Step1.} Initialize the cluster counter $Q'$ to $1$, assign all the $K$ sources to a single cluster $L_1$ and $\varnothing$ to a set of null clusters $A$.\\
{\bf Step2.} Choose cluster $L_{q^*}$, for $q^*\in\{1,\ldots,Q'\}-A$, with the largest angular diameter
\begin{equation}
D(L_{q^*})=\mbox{max}\{D(L_{q})|q\in\{1,\ldots,Q'\}-A\}.
\end{equation}
{\bf Step3.} Apply the presented weighted K-means clustering technique to split $L_{q^*}$ into two clusters, $L'_{q^*}$ and .$L''_{q^*}$\\\\
{\bf Step4.} If $D(L'_{q^*})+D(L''_{q^*})<D(L_{q^*})$, then set $Q'=Q'+1$, $L_{q^*}\equiv L'_{q^*}$, $L_{Q'}\equiv L''_{q^*}$, and $A=\varnothing$, otherwise set $A=A\cup\{q^*\}$.\\
Repeat steps 2, 3, and 4 until $Q'=Q$.

\section{Illustrative Example}
\label{Illustrative Example}
\begin{figure}
\vspace*{0.2cm}
\begin{center}
\hspace{-2mm}\epsfig{file=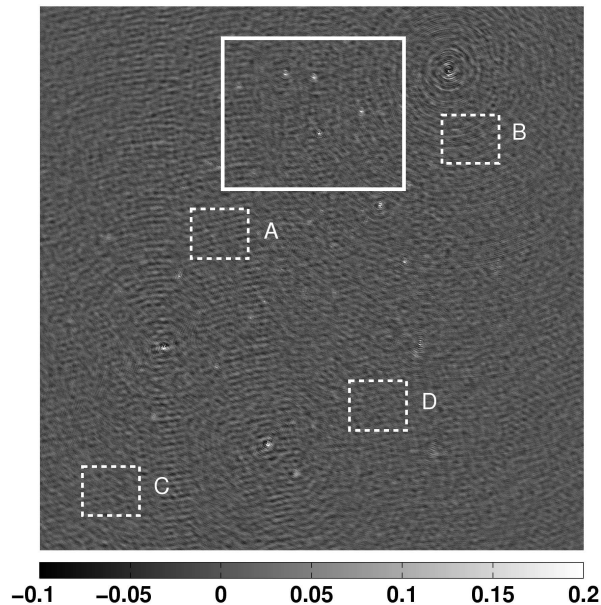, bb= 377 33 1127 788,clip=,scale=0.3} 
\vspace{-0.5cm}
\end{center}
\caption{ Ten averaged channels synthesis image of a 6 hour long  LOFAR 3C196 observation. The central source (3C196, peak flux is 70 Jy) plus the four brightest sources have been removed. Approximately 69 sources can be seen after the subtraction. The noise level is 6 mJy. }
\label{Figure1}
\end{figure}

\begin{figure}
\vspace*{0.05cm}
\begin{center}$
\begin{array}{cc}
\hspace{-4mm}\epsfig{file=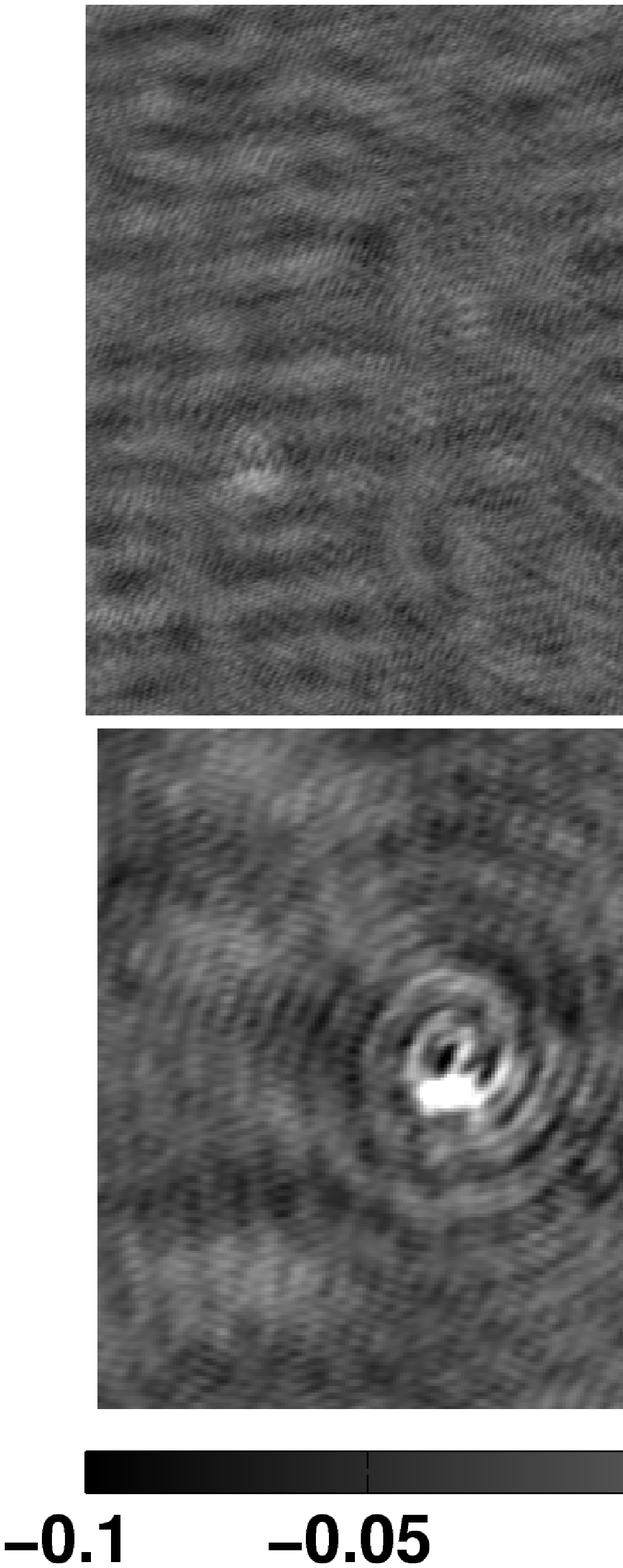,  bb= 305 18 1206 795,clip=,scale=0.14} &
\hspace{-3mm}\epsfig{file=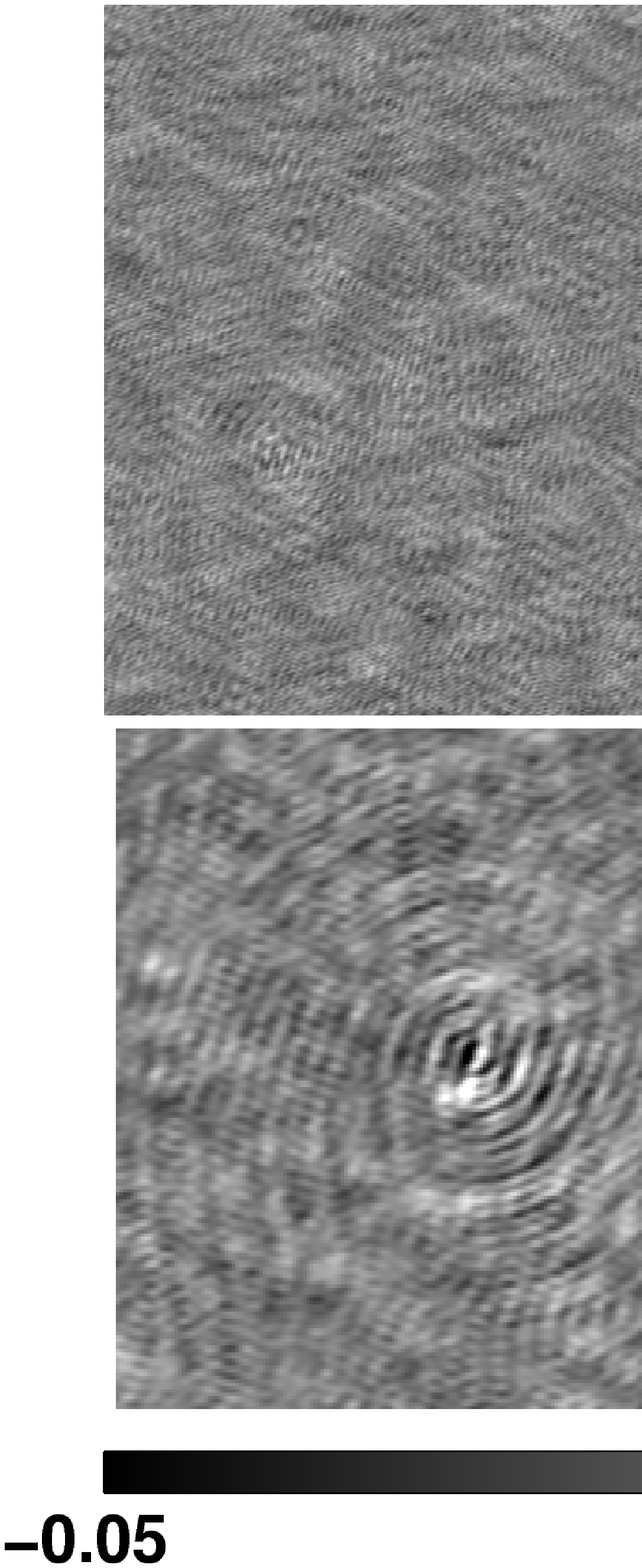,  bb= 308 18 1206 795,clip=,scale=0.14}\\[-1mm]
\hspace{-4mm}\epsfig{file=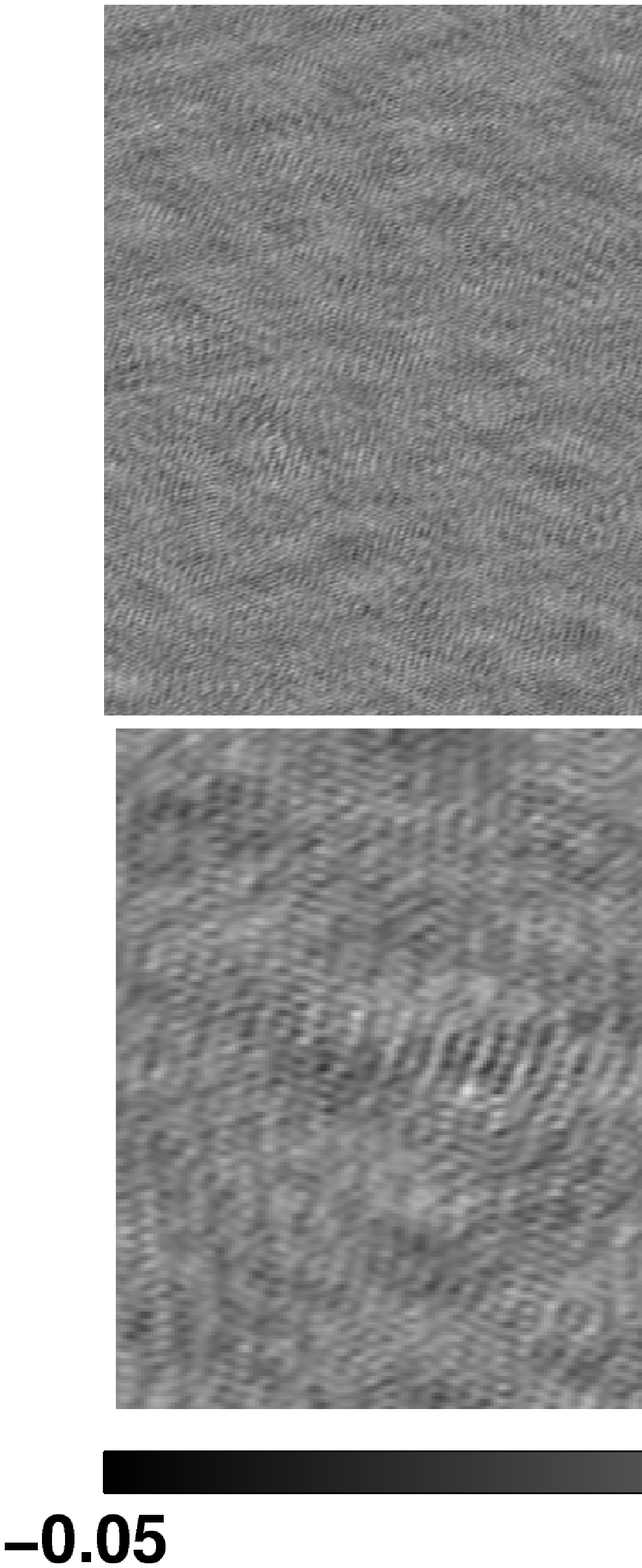,  bb= 308 18 1206 795,clip=,scale=0.14} &
\hspace{-3mm}\epsfig{file=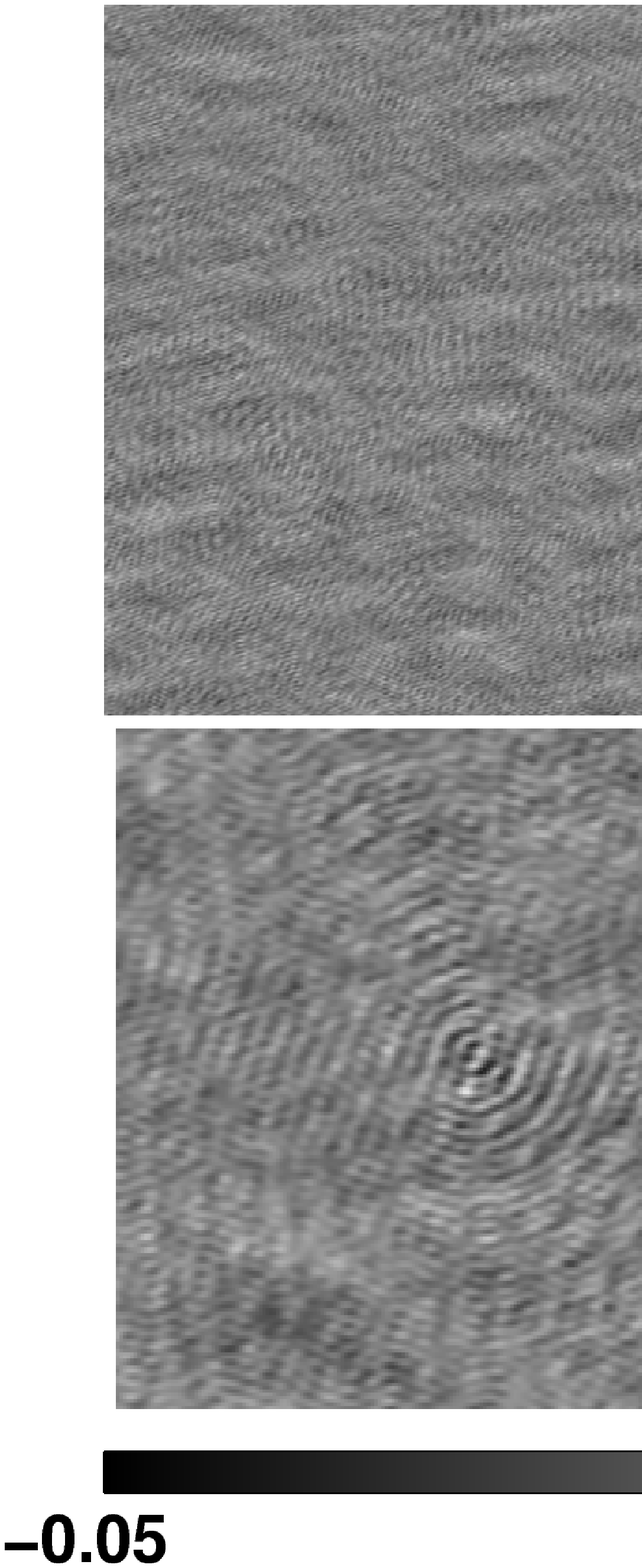,  bb= 308 18 1206 795,clip=,scale=0.14}
\end{array}$
\vspace{-.7cm}
\end{center}
\caption{ Zoomed in images obtained from the white window in Fig. \ref{Figure1}. The top row images are the initial image (left) and the residual image after subtracting 8 sources by the classical LS calibration (right). The residual images of the clustered calibration using Hierarchical (left) and Weighted K-means (right) clustering methods with 10 source clusters are shown at the bottom row. The inset figures show a blow-up of one of the 8 sources solved by the LS calibration.}
\label{Figure2}
\end{figure}

We consider the calibration of data obtained by LOFAR using 25 stations (receivers). The observation is centered at the radio source 3C196 and has an integration time of 6 hours. The central as well as the  four brightest sources were initially subtracted and the result is shown in Fig. \ref{Figure1}. We subsequently processed the same data using the classical calibration, in the direction of the 8 bright sources, and  the aforementioned  clustered calibration, with 10 clusters produced by weighted  K-means and divisive hierarchical clustering methods, on 30 seconds time intervals. For all the methods the LS optimization is used with 9 iterations and the residual images, zoomed into the area enclosed by the white window in Fig. \ref{Figure1}, are shown in Fig. \ref{Figure2}. The inset figures focusing on one of those 8 sources, randomly chosen, show that the source has been considerably better subtracted in the case of clustered calibration, while in the case of classical calibration  there is a significant residual error remaining.  Table \ref{table1} presents the comparison of the Root Mean Squared (RMS) of the residual maps for different regions and the full images produced by the classical and clustered calibration methods. The clustered calibration method has a lower RMS in all the cases.  

\begin{table}
\begin{center}
    \begin{tabular}{ c|c c c c c}
    RMS (mJy)& A & B & C & D & Full image\\
    \hline\\[-3mm]
    Classical & 6.5 & 6.0 & 5.6 & 6.5 & 6.8\\
    HC & 5.2 & 5.3 & 5.4 & 5.5 & 5.4\\
    WKC & 5.5 & 5.4 & 5.5 & 5.3 & 5.9\\
    \end{tabular}
\end{center}
\caption{The RMS of the residual images for the LS calibration without clustering (classical calibration), and by using  Hierarchical (HC) as well as Weighted K-means Clustering (WKC) of the sources. The letters A, B, C and D correspond to the regions demarcated by the boxes in Fig. \ref{Figure1}.  }  
\label{table1}
\end{table}

\section{Conclusions}
\label{Conclusions}
We have introduced a clustered calibration scheme for calibrating radio interferometric data towards the sensitivity limit. The method upgrades the coherencies of individual sources by their total amount obtained at each source cluster. Then, it applies calibration to these new coherencies that carry a higher level of information compared with the initial ones. Therefore, for calibration of sources bellow the noise level it has a considerably better performance compared with un-clustered calibration techniques. Divisive Hierarchical as well as Weighted K-means clustering methods are used to exploit the spatial proximity of the sources. It is also shown by an illustrative example that the RMS at different regions of the clustered calibration's residual images is consistently lower, when compared to the un-clustered calibration, which reveals its superiority at a low SNR. Hierarchical clustering provides a marginally better result since it constructs clusters of smaller angular diameters and thus it assigns the same calibration solutions to sources that have smaller angular separations. Future work will address the estimation of the optimum number of clusters, the performance of fuzzy clustering in the the clustered calibration, and combination of clustering with different calibration methods.

\bibliographystyle{IEEEbib}
\bibliography{references}

\end{document}